\documentclass[aps,pra, twocolumn, superscriptaddress,amssymb,showpacs]{revtex4}
\usepackage{graphicx}
\usepackage{epstopdf}

\begin{document}
\title{Chaos due to parametric excitation: phase space symmetry and photon correlations}

\author{T.~V.~Gevorgyan}
\email[]{t\_gevorgyan@ysu.am}
\affiliation{Institute for Physical Researches, National Academy
of Sciences,\\Ashtarak-2, 0203, Ashtarak, Armenia}

\author{G. H. Hovsepyan}
\email[]{gor.hovsepyan@ysu.am}
\affiliation{Institute for Physical Researches,
National Academy of Sciences,\\Ashtarak-2, 0203, Ashtarak,
Armenia}

\author{A.~R.~Shahinyan}
\email[]{anna\_shahinyan@ysu.am}
\affiliation{Institute for Physical Researches,
National Academy of Sciences,\\Ashtarak-2, 0203, Ashtarak, Armenia}

\author{G.~Yu.~Kryuchkyan}
\email[]{kryuchkyan@ysu.am}
\affiliation{Institute for Physical Researches,
National Academy of Sciences,\\Ashtarak-2, 0203, Ashtarak,
Armenia}\affiliation{ Yerevan State University, Centre of Quantum Technologies and New Materials, Alex Manoogian 1, 0025,
Yerevan, Armenia}

\begin{abstract}
{We discuss dissipative chaos showing symmetries in the phase space and
nonclassical statistics for a parametrically driven nonlinear Kerr resonator
(PDNR). In this system an oscillatory mode is created in the process of
degenerate down-conversion of photons under interaction with a train of
external Gaussian pulses. For chaotic regime we demonstrate, that the
Poincar\'e section showing a strange attractor, as well as the resonator mode contour plots of the Wigner functions display two-fold symmetry in the
phase space. We show that quantum-to-classical correspondence is strongly
violated for some chaotic regimes of the PDNR. Considering the second-order
correlation function we show that the high-level of photons correlation
leading to squeezing in the regular regime strongly decreases if the system
transits to the chaotic regime. Thus, observation of the photon-number
correlation allows to extract information about the chaotic regime.}
\end{abstract}

\maketitle

\section{Introduction}\label{introduction}

The nonlinear systems demonstrating chaotic behaviour are still the subject of
much attention. In this way many studies have been also made to explore quantum
effects in chaotic dynamics Refs. \cite {quant_chaos, quant_chaos1, quant_chaos2}.
Particularly, there has been a great interest in studying of driven quantum
chaotic systems in the presence of dissipation and decoherence, i. e.  quantum
chaos for open nonlinear systems. The beginning of study of an open chaotic
system can be dated back to the papers Refs. \cite{back, dit} where the authors have
analyzed the kicked rotor and similar systems with discrete time interacting
with a heat bath. At this point, we note that quite generally, chaos in
classically conservative and dissipative systems with noise, has completely
different properties. For example, strange attractors on Poincar\'e section can
appear only in dissipative systems. On the other hand, the manifestations of
chaotic motion in open quantum systems, though widely studied, remain somehow
not so clearly understood, both from the mathematical as well as from the
physical point of view. One reason is that by including dissipation and
decoherence, the computational effort grows drastically, since we have to
operate with density matrices instead of wave functions.

The most successful approach that probes quantum dissipative chaos seems to be
quantum tomographic methods based on the measurement of the Wigner function,
which is a quantum quasi-distribution in the phase space. In this way, the
connection between quantum and classical treatment of chaos can be realized by
means of a comparison between strange attractors in the classical Poincar\'e
section and the contour plots of the Wigner functions Refs. \cite{PhysE,mpop,Leon}.
On the other hand, alternative methods that probe quantum dissipative chaos
involve considerations of entropic characteristics, analysis of statistics of
excitation number Refs. \cite{K2,K3}, and the methods based on the fidelity decay
Refs. \cite{Lloyd,Weins,L2} and Kullback-Leibler quantum divergence Ref. \cite{Kulback},
and the purity of quantum states Ref. \cite{purOur}.

The simplest physical systems showing quantum dissipative chaos are based on
the model of kicked rotator. Its experimental realization, and observation of
the model’s dissipation and decoherence effects are carried out on a gas of
ultracold atoms in a magneto-optical trap subjected to a pulsed standing wave
Refs. \cite{Ammann41}, \cite{Klappauf42}. Recently, the nonlinear Kerr
resonator has been used for investigation of quantum chaos at a level of few
quanta Ref. \cite{purOur2}. 

The purpose of this paper is to investigate properties of dissipative chaos for
the other class of the simplest oscillatory open systems which possess some
remarkable properties, particularly, symmetries in the phase space as well as a
nonclassical statistics of excitation numbers. These symmetries can be
formulated on the system Hamiltonian and the master equation on one side or on
dynamical equations of motion in the semiclassical approach on the other side.
If the governing equations are invariant under a symmetric operation to some
state variables an interesting question is arises whether this symmetry
can be also seen in the chaotic regimes on the system attractors or the Wigner
functions. The answer of this question is very important for the deeper
understanding of quantum dissipative chaos.

The requirement in realization of this study is to have a proper quantum model
showing both chaotic dynamics, symmetry properties and nonclassical statistics
of excitation numbers. For this goal, in this paper we propose a parametrically
driven nonlinear Kerr resonator (PDNR) operated in the pulsed regime. In this
system an oscillatory mode is created in the process of degenerate
down-conversion of photons under an external Gaussian pulses.

The dynamics of periodically driven nonlinear oscillator has been also studied
in Ref. \cite{Dyk}.
Note, that an analogous system modeled as a kicked nonlinear system has been
proposed for investigation of quantum coherence and classical chaos Ref.
\cite{Milburn43}. It was also shown, that a more promising realization of this
system, including the quantum regime, is achieved in the dynamics of cooled and
trapped ions, interacting with a periodic sequence of both standing wave pulses
and Gaussian laser pulses Refs. \cite{Breslin44, Breslin45}. Recently the
experimental realization of quantum chaotic behavior for the model of quantum
kicked top on a single atom has been done Ref. \cite{chaos1}.  Transition to
classical chaos for a system with coupled internal (spin) and external
(motional) degrees of freedom has been also considered Refs. \cite{WW, WW1}.

The other important implementation of the PDNR is given by Josephson
junction embedded in a continuous linear circuit. Recently, it is consistently
demonstrated that the degenerate (as well as non-degenerate) parametric
interactions combined with strong Kerr-interaction can be realized in the
Josephson junction embedded in a continuous linear circuit Ref. \cite{Bour}. The
continuously driven version of the model was experimentally demonstrated for a
microwave cavity with a time-dependent boundary condition realizing by the
tunable inductance of a superconducting quantum interference device Ref.
\cite{Delsing}. Parametric conversion of microwave photons between two modes in
this Josephson parametric converter has already been observed Ref. \cite{Zak}.

It should be noted, that the combined parametrically driven anharmonic
oscillator under monochromatic driving has been proposed and studied in the
papers Refs. \cite{PNO1}, \cite{PNO2} for the special cases without
consideration of effects of dissipation. A complete quantum treatment  of
monochromatically driven parametric oscillator combined with Kerr-nonlinearity
has been developed in terms of the Fokker-Planck equation in complex
$P$-representation Refs. \cite{a33}, \cite{b33}. It has been demonstrated that
this system displays two-fold symmetry in the phase space. The cavity mode
displays two-phase stability: there are two stable states of mode with equal
photon numbers but with different phases, and as a result the Wigner function of
mode acquires two-fold symmetry in the phase space (see also Ref.
\cite{Tigran}). Note, that two-phase stability has been shown early for
monochromatically driven optical parametric oscillator above threshold of
generation Refs. \cite{Drum, Drum1, Drum2}. Below we show that such symmetry is
also displayed in the chaotic regime of PDAO under Gaussian pulses.

The other part of the paper is devoted to investigation of quantum-to-classical
correspondence as well as the photon-number correlation for oscillatory mode in
chaotic regime. We analyse phenomenon of photon correlation for regular and
chaotic regimes of the PDNR on the base of the second-order correlation function at
coinciding times that reflects the statistics of the oscillatory mode. In
general, the PDNR mode shows comparatively high-level of photon correlation due
to two-photon parametric excitation. As we will show below this level of photon
correlation decreases if the system is operated in the chaotic regime. Thus, we
conclude that by means of observing the correlator it will be possible to
extract information on the chaotic regime. Hence we present a simpler
manifestation of quantum chaos directly related to photon correlation. 

In the present paper the system under consideration may be modeled classically
and quantum mechanically with dissipation included in the both cases. The
classical dynamics of this system exhibits a rich structure of regular and
chaotic motion, with the parameters of the train of Gaussian pulses being the
control parameters. The quantum regime of the PDNR requires a comparatively
high level of third-order, Kerr nonlinearity with respect to dissipation, i.e.
$\chi/\gamma>1$, that is considered in this paper.

The paper is arranged as follows. In Sec.II we describe the PDNR under pulsed
excitation focusing on the phase space symmetry properties of the system. In
Sec. III, we study the properties of chaos on base of the Poincar\'e section,
the Wigner functions and the second-order correlation functions. We summarize
our results in Sec. IV.

\section{Pulsed the PDNR: phase-symmetry}\label{SecPD}

We start with the following Hamiltonian to describe
one-mode PDAO excited by the train of pulses
\begin{eqnarray}
H=\hbar \omega_{0}a^{+}a + \hbar \chi (a^{+}a)^{2}+~~~~~~~~~~~~~~~~~~~~~~\nonumber ~\\ \hbar f(t)({\Omega} e^{-i\omega t}a^{+2} + {\Omega^{*}}e^{i\omega t}a^{2})+H_{loss}.
\label{H}
\end{eqnarray}
Here, $a^{+}$, $a$ are the oscillatory creation and annihilation
operators, $ \omega_{0}$ is the oscillatory frequency, $ \omega$ is the mean frequency of the driving field and $\chi$ is
the nonlinearity strength proportional to the third-order
susceptibility for the case of Kerr-media. The coupling constant $\Omega f(t)$
is proportional to the second-order susceptibility and the time-dependent
amplitude of the driving field. It consists of the Gaussian pulses with 
duration $T$ which are separated by time intervals $\tau$
\begin{equation}
f(t)=\sum_{n=0}^{\infty}{e^{-(t - t_{0} - n\tau)^{2}/T^{2}}}.
\label{driving}
\end{equation}
$H_{loss}=a \Gamma^{+} + a^{+} \Gamma$ is responsible for the
linear losses of oscillatory mode, due to couplings with a heat
reservoir operators giving rise to the damping rate $\gamma$.

The reduced density operator of oscillatory mode $\rho$ within the framework
of the rotating-wave approximation, in the interaction picture corresponding
to the transformation $\rho \rightarrow e^{-i(\omega /2)a^{+}at} \rho
e^{i(\omega /2)a^{+}at}$ is governed by the master equation,

\begin{eqnarray}
\frac{d \rho}{dt} =-\frac{i}{\hbar}[H_{0}+H_{int}, \rho] + ~~~~~~~~~~~~~~~~~~~~~~~~\nonumber ~\\
\sum_{i=1,2}\left( L_{i}\rho
L_{i}^{+}-\frac{1}{2}L_{i}^{+}L_{i}\rho-\frac{1}{2}\rho L_{i}^{+}
L_{i}\right)\label{master},
\end{eqnarray}
where $L_{1}=\sqrt{(N+1)\gamma}a$ and $L_{2}=\sqrt{N\gamma}a^+$
are the Lindblad operators, $\gamma$ is a dissipation rate and N
denotes the mean number of quanta of a heat bath,

\begin{eqnarray}
H_{0}=\hbar \Delta a^{+}a, ~~~~~~~~~~~~~~~~~~~~~~~\nonumber ~\\
H_{int}= \hbar \chi (a^{+}a)^{2} + \hbar f(t)({\Omega}a^{+2} +
{\Omega^{*}}a^{2}),\label{hamiltonian4}
\end{eqnarray}
and $\Delta=\omega_{0} -\omega/2$ is the detuning between half
frequency of the driving field $ \omega/2$ and the oscillatory
frequency $ \omega_{0}$. To study pure quantum effects we focus on the cases
of very low reservoir temperatures for which the mean number of
reservoir photons $N=0$.

We use this system in the regime of strong Kerr nonlinearities with respect to
dissipation. Recent progress in circuit QED, superconducting systems and
solid-state artificial atoms has been opened up an effective additional path
for arranging devices describing by this model. Particularly, the same
Hamiltonian (\ref{H}) (with $f(t)=1$ ) describes the case of a Josephson
junction embedded in a transmission-line resonator in the case when the
effect of the quadratic part of the Josephson potential is taken into
account exactly. The $a^{+}$ and $a$ raising and lowering operators in this
case describe the normal mode of the resonator+junction circuit. The
nonlinearity is found to lead to self-Kerr effects. Then, by replacing the
single junction by a SQUID the Kerr coefficient allows the parametric
term describing degenerate two-photon excitation in microwave light
\cite{Bour}.

\subsection{Symmetry on the frame of Poincar\'e section}

In semiclassical approach the corresponding equation of motion
for the dimensionless amplitude of oscillatory mode has the following form

\begin{equation}
\frac{d\alpha}{dt}= -i[\Delta + \chi + 2|\alpha|^2\chi]\alpha + if(t)\Omega\alpha^{*} -\gamma\alpha.
\label{semclass}
\end{equation}

This equation modifies the standard equation for parametric oscillator with
Kerr nonlinearity in the case of pulsed excitation. It is used below for
comparison of the results obtained in quantum approach with analogous ones in
semiclassical limit.

Note that to calculate the Poincar\'e section, we have chosen  real $x_0$ and imaginary $y_0$ parts of the complex amplitude as an arbitrary initial phase space point of the system at the time $t_0$. We then
define a constant phase map in the $(X, Y)$ plane by the sequence of the periodically shifted points
$(X_n, Y_n)=(X(t_n), Y(t_n))$ at $t_n=t_0+n\tau$ ($n=0, 1, 2,...$). 
Poincar\'e section can be obtained based on $x=Re(\alpha)$ and $y=Im(\alpha)$.
It is evident from Eq. \ref {semclass} that the system has symmetry properties
in the phase space for $\alpha \rightarrow (-\alpha)$ replacement. It is obvious that this phase-symmetry 
is also displayed on the Poincar\'e section.

\subsection{Phase-space symmetry on the frame of Wigner functions}

It is evident that this symmetry is also  realized in quantum picture of the
the PDNR in the phase space. Really, considering the transformations:
\begin{equation}
H^{\prime}=U^{-1}HU,~~ \rho^{\prime}=U^{-1}\rho U
\label{Transform}
\end{equation}
with the unitary operator
\begin{equation}
U=exp\left(i \theta a^{+}a \right) \label{TransformOp}
\end{equation}
we verify that the interaction Hamiltonian (\ref{hamiltonian4})
satisfies the commutation relation
\begin{equation}
\left[H,U\right]=0, \label{ComRel}
\end{equation}
if the parameters $\theta$ are chosen as $\theta =\pi$.
The analogous symmetry takes place for the density operator of
oscillatory mode
\begin{equation}
\left[\rho(t),U\right]=0. \label{DisEf}
\end{equation}

One of the most important conclusions of such symmetries 
relates to the Wigner functions of the oscillatory mode
\begin{equation}
W\left(\alpha\right)=\frac{1}{\pi^{2}}\int d^{2}\gamma Tr\left(
\rho e^{\gamma a^{+}-\gamma^{*}a}\right) e^{\gamma^{*}\alpha
-\gamma \alpha^{*}}.
\label{Wig}
\end{equation}
In this formula, we perform rotations by the angle $\theta$ around the origin
in the phase spaces of complex variables $\alpha$ corresponding to the
field operators $a$ in the positive P-representation. Indeed, in
the polar coordinates $r,~\theta$ of the complex the phase space plane
$X=\left(\alpha + \alpha^{*}\right)/2=r\cos \theta$,
$Y=\left(\alpha - \alpha^{*}\right)/2i=r\sin \theta$ we derive
that the Wigner function displays two-fold symmetry in its
rotation around the origin of the phase space
\begin{equation}
W\left(r,\theta+\pi\right)=W\left(r,\theta \right).
\label{WigPol}
\end{equation}
It is well known that this symmetry takes place for the degenerate
optical parametric oscillator (OPO) reflecting on the
phase-locking phenomenon in above threshold regime of OPO.
According to phase-locking the mode of sub-harmonic generated in OPO are
produced with well-defined two phases Ref. \cite{Drum}. This situation takes place
also for the combined system under consideration due to the quadratic form of
nonlinear term in the Hamiltonian  Ref. \cite{b33}. We show below that two-fold symmetry is
also displayed  in the strange attractors for  the PDNR in the pulsed regime.

In the following the distribution of oscillatory excitation states
$P(n)=\langle n|\rho|n\rangle$, the  normalized second-order correlation function  $g^{(2)}$  as well as the Wigner functions
\begin{equation}
W(r, \theta)=\sum_{n,m}\rho_{nm}(t)W_{mn}(r,\theta)
\label{expr:wigner}
\end{equation}
in terms of the matrix elements $\rho_{nm}=\langle
n|\rho|m\rangle$ of the density operator in the Fock state
representation will be calculated. Here the coefficients
$W_{mn}(r,\theta)$ are the Fourier transform of the matrix elements of
the Wigner characteristic function.

We solve the master equation Eq. (\ref{master}) numerically based on the
quantum state diffusion method \cite{qsd}. The applications of this method for
studies of the driven nonlinear oscillators and OPOs can be found in Refs.
\cite{PhysE, mpop,  K2, K3,  qsch, Gev39}.

It should be mentioned that in the limit of very short duration of the pulses,
when Gaussian pulses in the formula Eq. \ref{driving} can be approximated by
$\delta(t - t_{0} - n\tau)$ functions, and for zero detuning the Hamiltonian
(\ref{hamiltonian4}) is coincided with the interaction Hamiltonian of the paper
Ref. \cite{Milburn43} devoted to study of parametrically driven anharmonic
oscillator. In cited paper a double peaked the phase space distribution of
oscillatory mode has been proposed. The Eq.(\ref{Wig}) of two-fold symmetry in the
phase space has verified this conjecture. We demonstrate below that this
phase-symmetry  of both  the classical equation and the master equation for the
chaotic dynamics leads to the symmetry of both  strange attractors and the
corresponding Winger functions.

\section{Peculiarities of chaotic regime for the PDNR}

\subsection{ Poincar\'e sections  and Wigner functions}

In this subsection we analyse chaos in the
framework of both  semiclassical and quantum probability  distributions by using a correspondence
between contour plots of the Wigner function and the Poincar\'e section. 
In semiclassical limit chaos is observed on the  Poincar\'e section that is
constructed  from the semiclassical  dimensionless position $x=Re(\alpha)$
and momentum $y=Im(\alpha)$ variables as solutions of  Eq. \ref{semclass} at
fixing points in the phase space at a sequence of periodic intervals. Choosing
$x_0$ and $y_0$ as an arbitrary initial the phase space point of the system at the
time $t_0$, we define a constant phase map in the plane by the sequence of
points at $t_n=t_0+n\tau$, where $n=0,1,2...$. This means that for any $t=t_n$
the system is at one of the points of the Poincar\'e section.  If the PDNR is
driven by the sequence of short pulses the dynamics of the system is
nonstationary hence the Poincar\'e section depends on the initial
time-interval $t_0$. We choose various initial time $t_0$ in order to ensure
that they match to the corresponding time-intervals of the Wigner function.

\begin{figure}
\includegraphics[width=8.6cm]{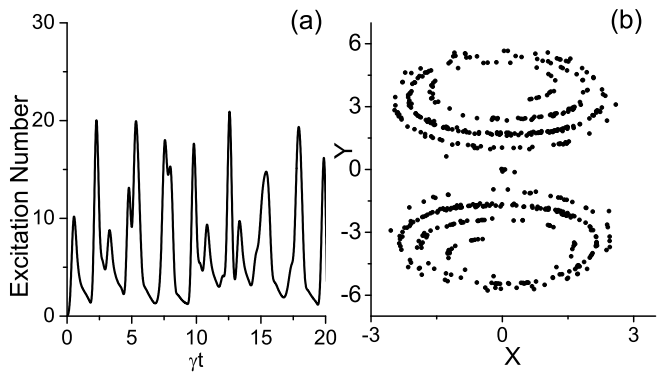}
\caption{Semiclassical results for the PDNR: the mean excitation numbers (photon numbers) versus dimensionless scaled time (a);   Poincar\'e section (b). The parameters are as follows:
$\Delta/\gamma = -20$, $\chi/\gamma=1$, $\Omega/\gamma=20$, (a) $T=0.5\gamma^{-1}$ $\tau=4\pi/5\gamma^{-1}$ .}
\label{fig.semPoincare}
\end{figure}

The typical results of calculations in the  semiclassical approach are depicted
in Fig. \ref{fig.semPoincare} for the mean excitation number  $|\alpha|^2 $ and
the Poincar\'e section  for the parameters $\Delta/\gamma $, $\chi/\gamma$,
$\Omega/\gamma$ as well as the parameters of pulses  corresponding to the
chaotic regimes. As we see, $|\alpha|^2 $ shows the usual chaotic dynamics
(see, Fig. \ref{fig.semPoincare}(a)) while   the  Poincar\'e section (see, Fig.
\ref{fig.semPoincare}(b)) displays  structure of the strange attractor. The
Poincar\'e section is symmetric over rotation on $\theta$ by the angle  around
the origin in the phase- spaces of complex variable for $\alpha \rightarrow
(-\alpha)$.

\begin{figure}
\includegraphics[width=8.6cm]{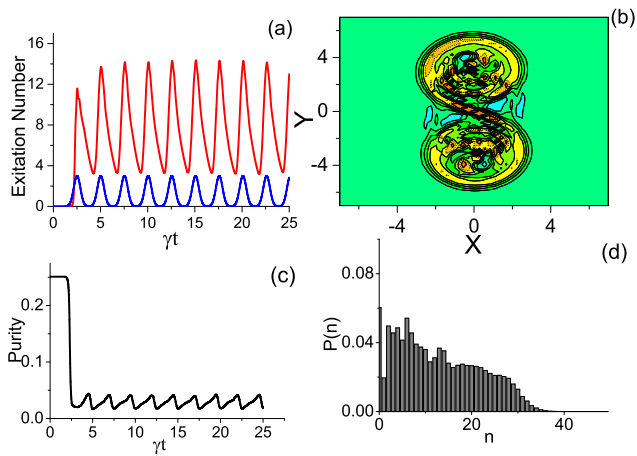}
\caption{(Color online) The mean excitation numbers of the PDNR obtained from  quantum averaging with
snapshots of Gaussian pulses that shows  below (a); contour plots of the Wigner function is the phase space (b); time-dependence of the quantum purity (c);  distribution of oscillatory excitation numbers (d). The parameters are
the same as for Fig. \ref{fig.semPoincare}:
$\Delta/\gamma = -20$, $\chi/\gamma=1$, $\Omega/\gamma=20$, (a) $T=0.5\gamma^{-1}$ $\tau=4\pi/5\gamma^{-1}$ .}
\label{fig.quantWig}
\end{figure}

Demonstration of chaos in the quantum treatment is presented in Fig.
(\ref{fig.quantWig}) by means of the mean excitation numbers (photon numbers), the contour plots
of the Wigner function, time-evolution of the quantum purity and the
distribution of oscillatory excitation numbers. Fig. \ref{fig.quantWig}(a) depicts the ensemble averaged mean oscillatory number. It is easy to see   that while the classical result (see, Fig. \ref{fig.semPoincare}(a)) shows the
usual chaotic behavior, its quantum ensemble counterpart (see,  Fig. (\ref{fig.quantWig}(a))  has clear regular behavior. These results indicate the well known result  that quantum dissipative
chaotic dynamics is not evident   in the mean oscillatory number and quantum
chaos can be displayed in the Wigner function.  In fact, it is easy to observe
in Fig. \ref{fig.quantWig}(b) that the contour plots of the Wigner function
are
generally similar to the corresponding classical  Poincar\'e section. Note that the ensemble-averaged mean oscillatory excitation
number and the Wigner functions are nonstationary and
exhibit a periodic time-dependent behavior, i.e., they repeat the
periodicity of the driving pulses at the overtransient regime.
We conclude that the Wigner function reflects the phase symmetry of the strange attractor.

Fig. \ref{fig.quantWig}(c) shows the distribution function of oscillatory excitation number 
$P(n)=\langle n|\rho|n\rangle$. It was demonstrated Ref. \cite{K2} that in transition from regular to chaotic dynamic of  a driven anharmonic oscillator  $P(n)$ is broadening. The analogous situation is realized for the PDNR. While $P(n)$ 
for regular dynamics  is clearly bell shaped and localized
in narrow intervals of oscillatory numbers, the distribution
for chaotic dynamics is flat topped with oscillatory numbers from $n=0$ to
$n_{max}$ 40 (see, Fig. \ref{fig.quantWig}(c)). The shape of distributions
changes irregularly in dependence from the duration $T$ and  time intervals
$\tau$ between pulses.

Correspondingly in Fig. \ref{fig.quantWig}(d) we plot quantum purity versus dimensionless scaled time intervals. 
This quantity has been used as a measure of the statistical characteristic of states and
decoherence.
The purity is defined through the density matrix $\rho$ of the system as
$P=Tr(\rho^2)$. This is measure of decoherence, for the pure state $Tr(\rho^2)
= 1$. As it is known that chaos affects coherence, so we can recognize chaos
within the framework of purity also.  Recently, the purity of quantum states has
been applied to probe chaotic dissipative dynamics.  

Analyzing the obtained results we can easily conclude that
system has chaotic dynamics in quantum limit also demonstrating the phase space
symmetry. These results are calculated
numerically on the base of master equation by means of the quantum state
diffusion method and for the same parameters of PDAO as it has been shown  in
Fig. \ref{fig.semPoincare}.

\subsection{ Violation of quantum-to-classical correspondence}

Another interesting part of our study is devoted to quantum-classical correspondence for the chaotic regime of the PDNR. 
 The connection between quantum and classical treatments of chaos is considered  by means
of comparison between strange attractors on the classical Poincar\'e  section
and the contour plots of the Wigner functions.

As shows results of calculations for the standard pulsed driven nonlinear oscillator Ref. \cite{PhysE},  for comparatively small values
of the ratio $\chi/\gamma$, , the contour plots of Wigner functions
are relatively close to the strange attractors. The main difference is that  the
Poincar$\acute{e}$ section has fine fractal structure while Wigner
function contour plot has not. This is due to the Hiesenberg
uncertainty relations which prevent sub-Plank structures in the phase space.
Such likeness of quantum and classical distribution vanishes in the deep
quantum regime.

\begin{figure}
\includegraphics[width=8.6cm]{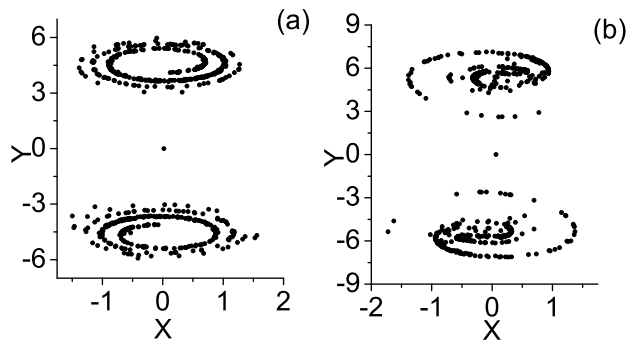}
\caption{Poincar\'e sections for the positive and negative detunings. The parameters are:
(a) $\Delta/\gamma = -10$, $\chi/\gamma=2$, $\Omega/\gamma=40$,$T=0.5\gamma^{-1}$ $\tau=\pi\gamma^{-1}$; (b) $\Delta/\gamma = 15$, $\chi/\gamma=1$, $\Omega/\gamma=10$, $T=0.5\gamma^{-1}$ $\tau=2\pi\gamma^{-1}$ .}
\label{fig.cl_poincares}
\end{figure}

As calculations show, the PDNR in the pulsed regime and in the semiclassical
approach displays chaotic dynamics  for both positive $\Delta>0$ and negative
$\Delta<0$ detunings. The corresponding Poincar\'e sections in the form of
strange attractors are depicted for both cases in Fig.
(\ref{fig.cl_poincares}). 
We note, that this result is exactly specific for the pulsed PDNR, as for a standard driven anharmonic oscillator in the  pulsed regime a chaotic regime is realized only for the case of negative detunings. 

\begin{figure}
\includegraphics[width=8.6cm]{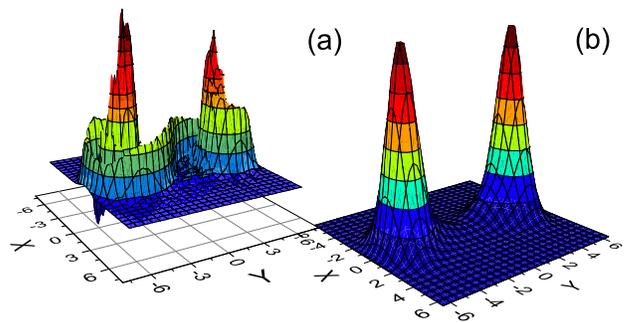}
\caption{(Color online) Wigner functions for the positive and negative detunings. The parameters are:
(a) $\Delta/\gamma = -7.5$, $\chi/\gamma=0.5$, $\Omega/\gamma=10$,$T=0.25\gamma^{-1}$ $\tau=2 \pi/5\gamma^{-1}$; (b) $\Delta/\gamma = 15$, $\chi/\gamma=1$, $\Omega/\gamma=10$, $T=0.5\gamma^{-1}$ $\tau=2\pi\gamma^{-1}$ .}
\label{fig.wigners}
\end{figure}

Analyzing the chaotic dynamics of the PDNR in quantum treatment  we conclude that quantum chaos is realized only for negative detunings, for  the parameters satisfies the following criteria:
$\Delta< 0$, $\Omega \simeq \vert\Delta \vert$, $\pi/2 \leq\tau/T \leq 2 \pi$,
in contrast to the results obtained in the semiclassical approximation. The
typical results for the Wigner functions corresponding to two cases of
detunings depicted in 
Figs. \ref{fig.wigners}. Comparing Poincar\'e section (see, Fig.
\ref{fig.cl_poincares}(b)) and the Wigner function (see,
Fig. \ref{fig.wigners}(b)) calculated for the same parameters with the positive
detuning we analyse quantum-classical correspondence. It is easy to conclude
that while in the semiclassical approach the PDNR shows chaotic regime in quantum
treatment the system displays regular dynamics. Indeed, the Wigner function
depicted in Fig. \ref{fig.wigners}(b) has two Gaussian-type peaks and displays a
twofold symmetry under the rotation of the phase space by angle $\pi$ around its
origin. For the case of negative detunings the Fig. \ref{fig.cl_poincares}(a)
clearly indicates the classical strong attractors with fractal structure that
are typical for a chaotic Poincar\'e section. The Wigner function Fig.
\ref{fig.wigners}(a) also reflects the chaotic dynamics, its contour plots in
the (x; y) plane are similar to the Poincar\'e section in form in contrast to
the case of positive detunings. It should be noted that the quantum-classical
correspondence is strongly  violated for some  regimes of the PDNR.

\subsection{Probe  of quantum chaos through photon-number correlations}
In this section we discuss the possibility of probing quantum chaos by observing photon statistics on base of the second-order correlation function  at coinciding times
\begin{equation}
g^{(2)}=\frac{\langle a^{\dagger}a^{\dagger}aa\rangle}{\langle a^{\dagger}a\rangle^2}.
\end{equation}

\begin{figure}
\includegraphics[width=8.6cm]{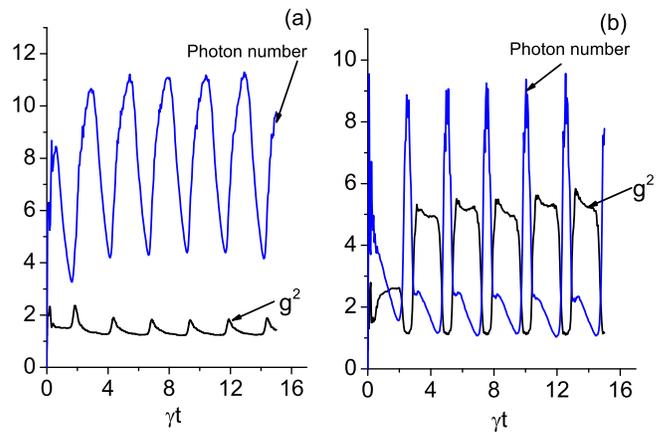}
\caption{The averaged photon numbers and the second-order correlation function for both chaotic and regular regimes of the PDNR. The parameters are: (a) $\Delta/\gamma = -20$, $\chi/\gamma=1$, $\Omega/\gamma=20$,  $T=0.5\gamma^{-1}$ $\tau=4\pi\gamma^{-1}/5$; (b) $\Delta/\gamma = 20$, $\chi/\gamma=1$, $\Omega/\gamma=20$,  $T=0.5\gamma^{-1}$ $\tau=4\pi/5\gamma^{-1}$.}
\label{fig.correlation}
\end{figure}

For this goal we calculate correlation function for both  regular and  chaotic regimes of the PDNR for two groups of the parameters leading to approximately equal excitation numbers for each of the  regimes. The correlation function is nonstationary and strongly depends on the properties of the input pulsed driving
field. The typical results for  $g^{(2)}$ in dependence of time-intervals are depicted in Figs. \ref{fig.correlation}.
\begin{figure}
\includegraphics[width=8.6cm]{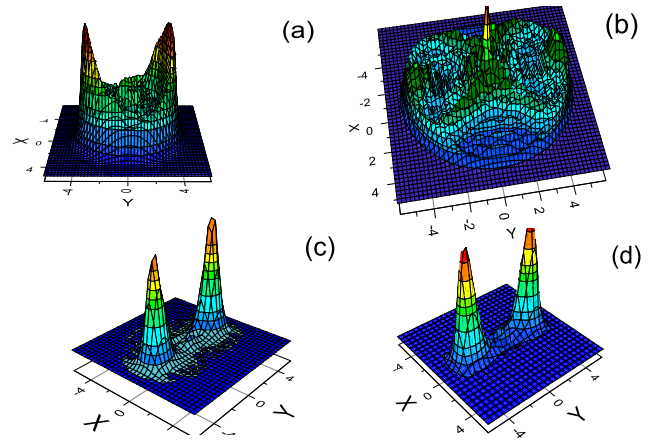}
\caption{The Wigner functions for time-intervals corresponding to the  minimum and
the maximum values of photon numbers depicted in Figs. \ref{fig.correlation} for the parameters: (a), (b) $\Delta/\gamma = -20$,
$\chi/\gamma=1$, $\Omega/\gamma=20$, $T=0.5\gamma^{-1}$
$\tau=4\pi\gamma^{-1}/5$;  (c), (d) $\Delta/\gamma = 20$, $\chi/\gamma=1$,
$\Omega/\gamma=20$,  $T=0.5\gamma^{-1}$ $\tau=4\pi/5\gamma^{-1}$. Figs. (a) and
(c) correspond to minimum values of photon numbers, (b) and (d) to the maximum values. 
 The Wigner functions on Figs. (a) and (b) show chaotic behaviour,  while  Fig. (c) and 
 (d) describe regular regime of the PDNR.}
\label{fig.minmax}
\end{figure}
Fig. \ref{fig.correlation}(a) corresponds to chaotic regime that is realized for the case of  negative detuning,  while Fig.
\ref{fig.correlation}(b) describes the  regular dynamic for positive detuning. As we see these results 
show drastically different behavior of $g^{(2)}$ for chaotic and regular regimes of the PDNR.  Let us consider this point in more details.

In the regular regime the minimal value of 
$g^{(2)}$ approximately equals to $1$ and  is realized for time intervals
corresponding to maximal values of the averaged photon numbers. In the
maximums the  correlation function describe photon bunching  that stipulated
by parametric two-photon processes leading to excitation of cavity mode.
Indeed, in this case   $g^{(2)}=5.2$ for the  mean photon number $ n=2.5$. To
illustrate this behaviour we also presentation on Fig. \ref{fig.minmax}(c, d)  the results on
the Wigner functions for the same parameters. As we see the results on
statistics of photons   are in accordance with  the Wigner functions. Indeed,
the Wigner functions showing the  regular regime  display two localized states
of the PDNR: for the case of  maximal value of photon number they  are closer to
coherent states, while for the case of photon bunching the peaks are squeezed
in the phase space.  For the chaotic regime the correlation function at the
maximum of photon number  approximately equals to  $2$ that indicates to
photon  statistics of thermal or chaotic light. For time intervals
corresponding to the minimum values of  photon numbers  $g^{(2)}$ is slightly
decreased.  The corresponding Wigner functions  depicted in Fig. \ref{fig.minmax} (a, b)
really confirm the chaotic regime of the PDNR. Thus, we conclude that two-photon
processes that stimulate  excitation of mode in the PDNR lead to strong
photon-number correlation in the regular regime,  but have not shown in the
chaotic regime. Nevertheless, this peculiarity of the PDNR is displayed as
two-fold symmetry in the phase space for both operational regimes.

\section{Summary}

We have shown pulse designed transition to chaos for the nonlinear dissipative
sytem. It has been demonstrated that nonlinear Kerr resonator parametrically
driven by the train of Gaussian pulses shows specific  chaotic behaviour, that
cardinally different from the analogous results obtained for  a standard
nonlinear Kerr resonator in the pulsed regime without parametrical
excitations.  We have shown that the Poincar\'e sections showing   strange
attractors in chaotic regimes as well as  the contour plots of the Wigner
functions of resonator mode  display  two-fold symmetry in the phase space.
This symmetry is naturally formulated as invariable of the system Hamiltonian
and the master equation under the symmetry operations. Therefore,  in this part
of the paper we have confirmed that  two-fold symmetry in the phase space is
conserved in transition of the PDNR to chaotic regime.  

The classical-to-quantum correspondence in chaotic regime  has been  considered
by means of comparison between strange attractors on the classical Poincar\'e
section and the contour plots of the Wigner functions. In this way, we have
demonstrated  that in the semiclassical approach  chaotic behavior of the PDNR
takes place  for both positive $\Delta>0$ and negative $\Delta<0$ detuning in
contrast to the case of standard nonlinear Kerr resonator. It has been shown
that  quantum-to-classical correspondence  in the phase space violates for some
regimes of the PDNR in the case of positive detunings $\Delta>0$. 

We have proposed to probe quantum chaos by observing of photon correlation  at
coinciding times intervals. In this way,  considering the second-order
correlation function of photon-number we demonstrate that the  high-level of
photon correlation leading to squeezing  in the regular regime strongly
decreases if the system transits to the chaotic regime. 

\begin{acknowledgments}
We acknowledge discussions with Lock Yue Chew and support from the Armenian State Committee of Science, the
Project No.13-1C031. 
\end{acknowledgments}

\end{document}